# Automating the RMF: Lessons from the FedRAMP 20x Pilot

Isaac Henry Teuscher
Brigham Young University
Provo, UT, USA
isaacteuscher@gmail.com

**ABSTRACT**

The U.S. Federal Risk and Authorization Management Program (FedRAMP®) has long relied on extensive sets of controls and static documentation to assess cloud systems. However, this manual, point-in-time approach has struggled to keep pace with cloud-native development. FedRAMP 20x, a 2025 pilot program, reimagines the NIST Risk Management Framework (RMF): replacing traditional NIST 800-53 controls with Key Security Indicators (KSIs), using automated, machine-readable evidence, and emphasizing continuous reporting and authorization.

This case study presents a practitioner-led field report from an industry participant who led multiple FedRAMP 20x pilot submissions and engaged directly with the FedRAMP PMO, 3PAOs, and community working groups. It explores how KSIs, continuous evidence pipelines, and DevSecOps integration can streamline authorization and improve cyber risk management. The study shows FedRAMP 20x as a live testbed for implementing the RMF in a cloud-native, automation-first approach and shares actionable recommendations for risk professionals seeking to modernize compliance and support real-time, risk-informed decision-making.

## 1 INTRODUCTION

### 1.1 Background on FedRAMP

The U.S. Federal Risk and Authorization Management Program (FedRAMP®) provides a standardized cybersecurity assessment and authorization process for cloud services [4]. Since 2011, FedRAMP has provided accreditation for cloud services (IaaS, PaaS, and SaaS) using security baselines derived from NIST SP 800-53 controls with a focus on cloud computing. This process involves a Cloud Service Provider's (CSP) attestation of control implementation in a written System Security Plan (SSP) document, an audit of the control implementation by an independent Third Party Assessment Organization (3PAO) and authorization by a U.S. Federal Agency (Agency). This process – having gone through five revisions – is currently known as Rev 5 Agency Authorization. The FedRAMP Rev 5 process is based on static documentation and annual assessments. A FedRAMP Rev 5 SSP often extends 800-1000+ pages and involves time intensive review and assessment. Due to the cost of time and resources required to undertake a FedRAMP Rev 5 Agency Authorization, in March 2025, a new FedRAMP process called FedRAMP 20x was launched aiming to overhaul FedRAMP to use automated, machine-readable evidence, and a set of Key Security Indicators (KSIs) instead of NIST 800-53 controls [3]. FedRAMP 20x is being developed "entirely in public" together "with industry" partners who provide feedback and participate in pilot phases to test and iterate on the new requirements [5].

### 1.2 Background on NIST RMF

In 2010, the U.S. National Institute of Science and Technology (NIST) developed and released what would become the Risk Management Framework (RMF) to provide guidance to U.S. federal agencies and other organizations seeking to manage cyber risk and comply with mandates from the Federal Information Security Modernization Act (FISMA) of 2014 (an update to the Federal Information Security Act of 2002). The NIST RMF has become a well-known and frequently used approach to cyber risk management. In the current second revision, the RMF includes seven steps to plan, develop, apply, and report on an organization's risk posture. These steps are: 1) Prepare, 2) Categorize, 3) Select, 4) Implement, 5) Assess, 6) Authorize, and 7) Monitor [1].

### 1.3 Research Premise

Since its inception, FedRAMP has used the NIST Risk Management Framework (RMF) process and static control documentation to assess cloud systems. This manual, point-in-time approach has struggled to keep pace with modern cloud-native development. The traditional Rev 5 process is misaligned with the speed, architecture, and dynamic nature of cloud-native systems, creating lagging indicators of cyber risk posture. FedRAMP 20x, the new pilot authorization path, represents a significant shift to informing risk decisions in real time. FedRAMP 20x is a testbed for modernizing the Risk Management Framework using machine-readable evidence, continuous risk metrics, and quantifiable control assurance. Due to the scope of U.S. federal cloud-service use, FedRAMP has become a complex and important process in U.S. enterprise cybersecurity risk management. FedRAMP 20x offers a valuable case study for modernizing enterprise risk management and performing cyber risk assessment through automation and continuous evidence.

This research paper presents a field study from the perspective of a participant observer (an industry partner undertaking the 20x pilot, not government staff) that explores how FedRAMP 20x reimagines the RMF implementation. The author contributed frequently to FedRAMP 20x Community Working Group discussions and interacted with the FedRAMP PMO, accredited 3PAOs, and leading GRC vendors to understand the goals and experience of FedRAMP 20x pilot participants in addition to leading the 20x assessment process and submission for multiple cloud-native products. Prior to engaging with the 20x pilot, the author participated in a Rev 5 FedRAMP assessment and worked as a security engineer for a GRC vendor that provides FedRAMP compliance related software.

Lessons learned from building an enterprise risk management process to support FedRAMP 20x requirements are shared providing actionable lessons for risk professionals seeking to modernize compliance, enhance measurement of cyber risk, and integrate automated governance into enterprise risk programs.

## 2 FEDRAMP 20X REQUIREMENTS

### 2.1 Participation

The case subject of this research is the FedRAMP 20x Phase One Pilot program undertaken in the Spring-Summer of 2025. Organizations with cloud-native software as a service (SaaS) products with existing cybersecurity compliance work accomplished (such as a SOC 2 Type 2 audit) and resources to develop solutions to meet the FedRAMP 20x requirements were encouraged to participate.

### 2.2 Key Security Indicators (KSIs)

FedRAMP 20x requirements center around Key Security Indicators (KSI) developed for 20x by the FedRAMP PMO. A KSI is a security capability that must be in place for FedRAMP approval. KSIs are specific to modern cloud-native architectures and focus on critical security safeguards. KSIs are designed to provide a standardized, machine-readable way to evaluate and monitor the security of a CSP. They focus on measurable outcomes rather than prescriptive processes to give CSPs more flexibility in how they demonstrate compliance.

KSIs are not directly tied or mapped to NIST 800-53 controls; however, the NIST control catalog served as a basis for the development of the 20x KSIs. KSIs are based on Rev 5 FedRAMP baseline controls but modernized for cloud native offerings and condensed to focus on key capabilities. Each KSI includes several validations or sub-parts that give specific direction on how to implement and assess each KSI. In this way KSIs function like NIST control families (e.g.: Access Control AC) and controls (AC-1). In total there are seven KSIs with 51 KSI validations [6]. A sample of KSIs are shown in Table 1.

| KSI ID | KSI Validations |
|---|---|
| KSI-CNA-01 | Configure ALL information resources to limit inbound and outbound traffic |
| KSI-CNA-02 | Design systems to minimize the attack surface and minimize lateral movement if compromised |
| KSI-CNA-03 | Use logical networking and related capabilities to enforce traffic flow controls |
| KSI-CNA-04 | Use immutable infrastructure with strictly defined functionality and privileges by default |
| KSI-CNA-05 | Have denial of service protection |
| KSI-CNA-06 | Design systems for high availability and rapid recovery |
| KSI-CNA-07 | Ensure cloud-native information resources are implemented based on host provider's best practices and documented guidance |

**Table 1. Cloud Native Architecture KSI**

Due to the cloud-native nature of 20x, no KSIs are designed to be fully inherited by a Cloud Service Provider (CSP). Unlike the NIST 800-53 catalog where a cloud-native product would fully inherit certain controls (e.g., Physical and Environmental Security) with KSIs there is a responsibility for the CSP to configure and demonstrate for each KSI validation. This is evident by the NIST 800-53 "Related Controls" list provided by the PMO with the initial set of KSIs fully excluding controls from the Maintenance (MA), Media Protection (MP), and Physical and Environmental Protection (PE) control families that are included in the FedRAMP Rev 5 Low Baseline [6].

### 2.3 Automation

One of the driving goals of FedRAMP 20x is to maximize automation so that "80%+ of requirements will have automated validation" [5]. During the pilot phase CSPs were encouraged to automate evidence collection and assessment of KSIs as much as possible with the understanding that some KSIs would remain as manual, point-in-time check for the time being.

## 3  FINDINGS

Several lessons were learned through the process of building an enterprise risk management process to support the FedRAMP 20x requirements. These findings are summarized below and presented with references to the NIST RMF to illustrate how FedRAMP 20x relates to known risk management practices.

### 3.1 FedRAMP 20x vs FedRAMP Rev 5 Process

When analyzing the FedRAMP 20x experience compared to a FedRAMP Rev 5 process through the lens of the Risk Management Framework, a few notable improvements are found. The RMF Selection step is faster due to a reduced number of KSIs compared to the traditional control baselines and cloud-native inheritance. The RMF Implementation step is driven by the system architecture, much of which can be gathered from a Cloud Security Posture Management (CSPM) tool or features provided by a cloud hosting provider (e.g., AWS, Azure, GCP). Major improvements are made at the Assessment and Authorization RMF steps. Evidence for 20x is mostly gathered through APIs and automated processes and reported through live dashboards. Assessment packages are changed from extensive written narratives (e.g., SSP, SAR) to a machine-readable file. Two significant challenges with Rev 5 were the cost of time and resources to write the SSP and to manually assess each control implementation. The 20x process offers an alternative to both challenges, reducing time to authorization and cost of compliance. Lastly, the RMF Monitoring step shows improvement as 20x enables real-time, continuous reporting instead of periodic, point-in-time reporting where the actual state of implementation may not accurately be reflected in the system's security documentation.

### 3.2 Lessons Learned

#### 3.2.1 Maximize Use of SDLC

If an organization has a thoughtfully designed system development life cycle (SDLC) process for example: a CI/CD pipeline with security testing steps, a Change Management process, using infrastructure as code, that system is well on its way to performing effective risk management and likely covers many KSIs already.

> "The best RMF implementation is one that is indistinguishable from the routine SDLC processes carried out by organizations. That is, RMF tasks are closely aligned with the ongoing activities in the SDLC processes, ensuring the seamless integration of security and privacy protections into organizational systems—and taking maximum advantage of the artifacts generated by the SDLC processes to produce the necessary evidence in authorization packages to facilitate credible, risk-based decision making..." - NIST SP 800-37 r2, page 23

- "The artifacts generated by the SDLC processes" — For example, a Terraform state file, a JSON output from a CI/CD pipeline or change control board approval.
- "Produce the necessary evidence" — Automated, machine-readable and ongoing rather than screenshots or point-in-time samples.
- "In authorization packages" — The 20x machine-readable assessment file which replaces the SSP
- "To facilitate credible, risk-based decision making" — This is precisely what FedRAMP 20x is all about

#### 3.2.2 Shift Left

The DevSecOps concept of "shift left" applies to risk management and FedRAMP compliance similar to how it applies in the software development process. As an organization makes smart choices in the design and gap assessment phase there are significant savings during the audit and continuous monitoring periods. 20x rewards early architecture decisions. For example, we use a vulnerability scanning tool that typically runs on a virtual machine or server, but we wanted our FedRAMP environment to only use containers with no managed operating systems (to reduce our attack surface, complexity, and operating system vulnerability management). We designed a way to deploy the scanning tool in Kubernetes. By making that decision at the architecture design step, we eliminated a lot of security compliance work later. Time deciding what should be in and out of the FedRAMP environment early in the process is time well spent.

> "Without the early integration of security and privacy requirements, significant expense may be incurred by the organization later in the life cycle to address security and privacy concerns that could have been included in the initial design. When security and privacy requirements are defined early in the SDLC and integrated with other system requirements, the resulting system has fewer deficiencies, and therefore, fewer privacy risks or security vulnerabilities that can be exploited in the future… Risk management activities begin early in the SDLC and continue throughout the life cycle." - NIST SP 800-37 r2, page 162

#### 3.2.3 Maximize Use of Automation

To support the goal of FedRAMP 20x to have "80%+ of requirements with automated validation" [5], our 20x process and pilot submission involved minimal staff (1-2 full-time engineers) since we leveraged AI tools and built off existing risk assessment work done (prior audits and existing security posture).

> "Organizations should maximize the use of automation, wherever possible, to increase the speed, effectiveness, and efficiency of executing the steps in the Risk Management Framework (RMF). Automation is particularly useful in the assessment and continuous monitoring of controls, [and] the preparation of authorization packages… Organizations have significant flexibility in deciding when, where, and how to use automation or automated support tools for their security and privacy programs." - NIST SP 800-37 r2, page xii

An in-depth review of how to effectively automate the testing of controls is given in NIST IR 8011v1r1 IPD Testable Controls and Security Capabilities for Continuous Monitoring Volume 1 [2].

#### 3.2.4. Maximize Use of Common Controls

FedRAMP 20x encouraged organizations who had completed a recent SOC 2 Type 2 or similar audit to adapt that evidence and tooling for FedRAMP 20x. In our experience, we were able to align many of the same capabilities and evidence produced for previous audits to the FedRAMP 20x KSIs. NIST SP 800-37 r2 recommend the following:

> "Maximize the use of common controls to promote standardized, consistent, and cost-effective security and privacy capability inheritance.… Maximize the use of automated tools to manage security categorization; control selection, assessment, and monitoring; and the authorization process." - NIST SP 800-37 r2, page 25

This has been extremely successful for us. We use common controls to maximize inheritance and maintain consistent implementations across 800-53 and 800-171 control sets (FedRAMP Rev 5, CMMC, SOC 2, and other compliance frameworks) and the 20x KSIs. We use a GRC tool to manage assessment and monitoring.

#### 3.2.5. Aim for Continuous Risk Assessment, Not Perfection

While some wonder if a partial status on a KSI will cause their product to "fail" FedRAMP 20x, the goal is an accurate, continuous risk assessment, not a perfect score. If the risk assessment is accurate and ongoing, the 20x process is working properly. Risk-based decisions need to be made by agencies who use cloud offering and need to be made on an ongoing basis by each CSP.

> "Respond to risk based on the results of ongoing monitoring activities, risk assessments, and outstanding items in plans of action and milestones."
> - NIST SP 800-37 r2, page 78

It's important to remember the goal is not simply to report on status but to enable risk-based decision making. NIST IR 8011v1r1 IPD states this finding clearly:

> "The objective of continuous monitoring is to respond to identified risks, not just to display results or generate reports"
> - NIST IR 8011v1r1 IPD, page 40

## 4  CONCLUSION

The lessons learned from participating in the FedRAMP 20x pilot provide actionable suggestions for Cloud Providers and Risk Analysts. For organizations seeking to implement a 20x approach to risk management it is critical to integrate evidence collection into DevSecOps pipelines and stay focused on cloud-native security best practices. For risk analysts, the 20x KSIs present a new framework that can be mapped to threat models or risk factors to contribute to holistic enterprise risk management. Additionally, the automated evidence collection, assessment, and reporting proposed by FedRAMP 20x presents an opportunity to automatically update and quantify residual risk beyond the simple metric of KSI implementation status. Ultimately, FedRAMP 20x demonstrates that the NIST RMF can evolve to meet the needs of modern cloud-native environments and provide a thorough authorization process with reduced manual compliance burden. This field study of the FedRAMP 20x Phase One Pilot presents one perspective into a broader shift toward continuous authorization, automated evidence, and better risk-informed governance.